\documentclass[fleqn,twoside]{article}
\usepackage{espcrc2}

\usepackage{graphicx}
\usepackage[figuresright]{rotating}

\newcommand{\AmS}{{\protect\the\textfont2
  A\kern-.1667em\lower.5ex\hbox{M}\kern-.125emS}}

\title{Search for Small-Scale Anisotropy of Cosmic Rays above $10^{19}$\,eV
       with HiRes Stereo}

\author{Stefan Westerhoff\address[MCSD]
       {Columbia University, Department of Physics,\\
        538 West 120 Street, New York, NY 10027, USA\\
        westerhoff@nevis.columbia.edu}
        for the High Resolution Fly's Eye (HiRes) 
        Collaboration
        \thanks{See http://hires.phys.columbia.edu for a complete list of authors}}
       
\begin{document}

\begin{abstract}
We present the results of a search for small-scale anisotropy in the 
distribution of arrival directions of cosmic rays above $10^{19}$\,eV 
measured in stereo by the High Resolution Fly's Eye (HiRes) experiment.
Performing an autocorrelation scan in energy and angular separation,
we find that the strongest correlation signal in the HiRes stereo data set
recorded between December 1999 and January 2004 is consistent with the 
null hypothesis of isotropically distributed arrival directions.  These 
results are compared to previous claims of significant small-scale 
clustering in the AGASA data set.
\vspace{1pc}
\end{abstract}

\maketitle

\section{Introduction}

Among the most striking astrophysical phenomena today is the
existence of cosmic ray particles with energies in excess of
$10^{20}$\,eV.  While their presence has been confirmed by a number
of experiments, it is not clear where and how these particles
are accelerated to these energies and how they travel astronomical
distances without substantial energy loss.

At energies above several $10^{18}$\,eV, cosmic ray particles
are believed to be of extragalactic origin.  If charged cosmic 
ray particles do not suffer considerable deflections in Galactic 
and extragalactic magnetic fields, one can hope to identify the 
sources and understand the underlying acceleration mechanism by a 
detailed study of their arrival directions.  The strength and
orientation of these fields is poorly known and estimates 
vary~\cite{sigl2003,dolag2003}, but their impact should decrease
at the largest energies; here, cosmic ray astronomy might be possible.

The small cosmic ray data set has been subjected to extensive 
searches for an autocorrelation signal, {\it i.e.} clustering 
of arrival directions on small angular scales, and there have been 
a variety of attempts to correlate catalogs of known astrophysical 
sources with cosmic ray arrival directions. 

So far, all efforts to identify the sources from the sparsely 
populated skymap have not produced statistically convincing
evidence for autocorrelation or correlations 
with any class of objects.  ``Statistically convincing'' should be
emphasized here, as there is actually no shortage of claims for
both clustering and correlation with catalogs.
Small-scale clustering of ultrahigh energy cosmic rays above 
$4\times 10^{19}$\,eV has, for example, been repeatedly
reported~\cite{agasa1996,agasa1999,agasa2001,tinyakov2001,agasa2003}, 
with analyses mainly based on arrival directions recorded with the 
Akeno Giant Air Shower Array (AGASA) in Japan.  If correct, these
results could indicate that cosmic rays originate in nearby, compact
sources.

\begin{figure}[ht]
\includegraphics*[width=19pc]{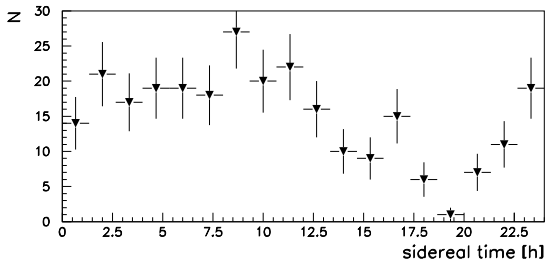}
\caption{\it Sidereal times of HiRes stereo events with energies 
$E>10$\,EeV.
\label{sidereal}}
\end{figure}

However, there is considerable disagreement over the statistical 
significance of this clustering signal.  The problem arises from the way 
the chance probability of the signal is evaluated.  Quite often, the data 
set used to formulate the correlation hypothesis is also used for
evaluating its significance.  Problems with published claims of 
significant small-scale clustering have been pointed out 
by various authors~\cite{watson2001,evans2003,finley2004}, and it has 
become clear that ultimately, only statistically independent data sets 
will allow a rigorous test of these claims.

Such a data set has recently become available.  The High Resolution 
Fly's Eye (HiRes) experiment in Utah has operated in stereo mode 
since December 1999.  The operation of HiRes stereo and AGASA overlaps 
only a few months in time, but both detectors are located in the northern 
hemisphere and probe approximately the same part of the northern sky.  
The HiRes stereo data set therefore provides an opportunity to independently 
test previous claims that the arrival direction of ultrahigh energy cosmic
rays shows statistically significant small-scale clustering.

\section{HiRes Stereo}

\begin{figure}[ht]
\includegraphics*[width=19pc]{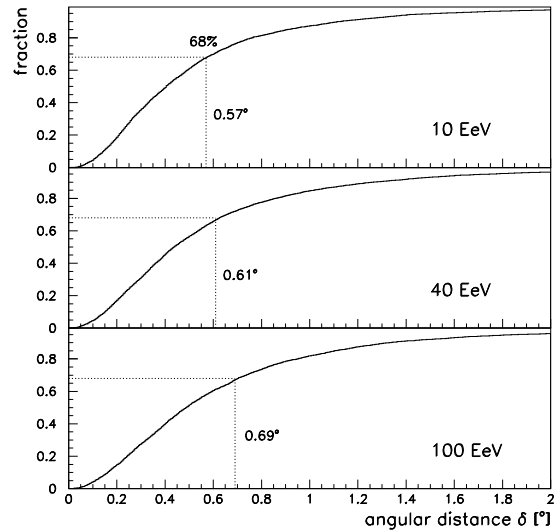}
\caption{\it Angular resolution of HiRes stereo for energies 10, 40, and
100 EeV.  The plot shows the fraction of simulated events reconstructed 
within $\delta$ degrees of the true arrival direction, as a function of 
$\delta$.
\label{fraction}}
\end{figure}

\begin{figure}[ht]
\includegraphics*[width=19pc]{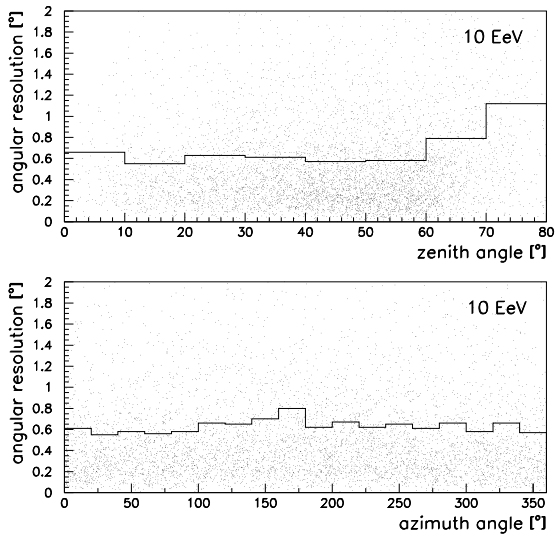}
\caption{\it Angular resolution as a function of zenith angle (top)
and azimuth angle (bottom) for simulated showers with energy $E=10$\,EeV.
The scatter plot shows individual events, the solid line represents
the angular distance which includes 68\,\% of all events in the 
zenith or azimuth angle bin.
\label{angular}}
\end{figure}

HiRes is an air fluorescence experiment with two sites
(HiRes\,1 and 2) at the US Army Dugway Proving Ground in the
Utah desert ($112^{\circ}$\,W longitude, $40^{\circ}$\,N latitude,
vertical atmospheric depth $860\,{\mathrm{g}}/{\mathrm{cm}}^{2}$).
The two sites are separated by a distance of 12.6\,km.
A general description of the HiRes detector is given
in~\cite{thomson2004}.  

Air fluorescence detectors observe air showers from the fluorescence 
light emitted by air molecules excited by particles 
in the shower.  Tracks of air showers developing in the atmosphere are 
directly viewed by photomultiplier cameras that continuously watch 
the night sky.

With showers detected out to distances of about 60\,km, the
instantaneous aperture of an air fluorescence telescope is quite
large.  However, the operating conditions of such an instrument
-- dark, moonless nights with a clear atmosphere -- restrict its
duty cycle to about $10\,\%$.  Moreover, the sidereal time distribution 
(Fig.\,\ref{sidereal}) is modulated by overall seasonal variations as 
a result of the extended observation time during the winter months. 

In monocular reconstruction, the so-called shower detector plane, which 
is the plane that contains the shower track and the detector, is very 
well-defined, whereas the location of the shower within that plane has 
to be reconstructed from the trigger times of individual phototubes 
along the shower track.  This leads to fairly large errors, of order 
several degrees, on the angle describing the shower in the plane.  
The ambiguity is resolved if the shower is seen by two detectors.  
Stereo data has the distinct advantage that error bars on the arrival 
direction are fairly symmetric for most azimuth angles.  A global 
$\chi^{2}$-minimization using all available information, including 
the tube trigger times, gives an angular resolution which is typically 
about half a degree.

The angular resolution of HiRes is determined using simulated showers.
We use a full detector simulation of proton showers generated with 
CORSIKA 6 \cite{corsika1998} using QGSJET for the first interaction.  
The showers are thrown isotropically and undergo a full detector
simulation and event reconstruction including all cuts that are
applied to the real data~\cite{apjl2004}.  As shown in 
Fig.\,\ref{fraction},  68\,\% of all showers generated at $10^{19}$\,eV 
are reconstructed within less than $0.57^{\circ}$ of the true shower 
direction.  The angular resolution depends weakly on energy;
the 68\,\% error radius grows to $0.61^{\circ}$ and $0.69^{\circ}$ for 
showers generated at $4\cdot10^{19}$\,eV and $10^{20}$\,eV, respectively, 
because at higher energies, showers are on average farther away.

Fig.\,\ref{angular} shows the angular distance between true and 
reconstructed shower direction for a large number of showers with 
energy $E=10^{19}$\,eV as a function of zenith and azimuth angle 
of the arrival direction.
The angular resolution is essentially constant in zenith 
and azimuth angle, varying by less than $0.1^{\circ}$ for zenith 
angles less than $70^{\circ}$.   There is a small range of azimuthal 
angles where the angular reconstruction is notably worse: for azimuth 
angles where the shower and the two sites are in the {\it same} 
plane, the stereo reconstruction looses its advantage and is essentially 
reduced to a monocular reconstruction.  Showers having poor angular 
resolution as a result of this ambiguity fail the quality cuts.
 
The major systematic errors on the arrival directions are due to 
uncertainties in the mirror pointing directions.  Reconstruction of laser
tracks and observation of UV bright stars in the field of view of
the cameras~\cite{sadowski2002} indicate that this error is not 
larger than $0.2^{\circ}$.

No explicit weather cut is applied for this data set.  Simulations show
that as long as weather conditions are known, their impact on the 
determination of the shower geometry is minimal.  The reconstruction
uses an hourly atmospheric data base built from reconstructed laser shots.
Note that stereo data provides us with a consistency check regarding the
weather corrections applied to the data.  If the distance of the shower 
to the two detectors is notably different, any inaccurate correction 
for light scattering and absorption will result in a mismatch of energy 
estimates for the two sites.

\begin{figure*}[ht]
\includegraphics*[width=35pc]{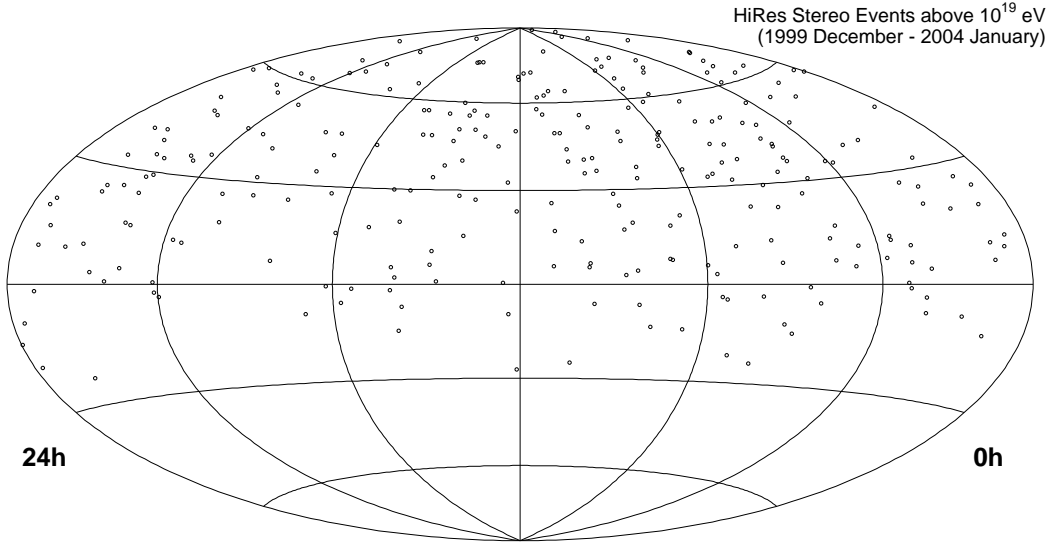}
\caption{\it Arrival directions in equatorial coordinates for
271 HiRes stereo events above 10\,EeV recorded between
December 1999 and January 2004.
\label{skymap}}
\end{figure*}

After quality cuts~\cite{apjl2004}, the stereo data set used in this
analysis comprises 271 events above $10^{19}$\,eV recorded between 
December 1999 and January 2004.  Fig.\,\ref{skymap} shows a skymap 
of their arrival directions in equatorial coordinates.

\section{Small-Scale Clustering}

As described in~\cite{apjl2004}, we search for small-scale clustering
by performing an autocorrelation scan in energy and angular
separation.  For a given energy $E$, we count the number of
pairs $n_{p}$ separated by less than $\theta$.  We then use
sets of simulated showers with the same number of events as the HiRes
stereo data set, but with an isotropic distribution of arrival
directions, to estimate the chance probability $P(E,\theta)$
of finding $n_{p}$ or more pairs in a random data set just by chance.  
We identify the energy threshold $E_{min}$ and the angular distance 
$\theta_{min}$ which gives the smallest chance probability, $P_{min}$.
This is the most promising potential clustering signal, and its
statistical significance can be determined by performing the 
same scan over $n_{MC}$ data sets with isotropic arrival directions,
finding the minimum probability for each of these sets, and
counting the number of data sets for which $P<P_{min}$.
Simulations show~\cite{apjl2004} that despite the statistical penalty 
incurred by scanning, this procedure results in a final 
chance probability of about $1\,\%$ ($2\times 10^{-3}\,\%$) for as
few as 3 (4) clusters in 47 events, assuming that cosmic rays are
not subject to strong deflections in magnetic fields.

For the HiRes stereo data set, we perform this scan over the
271 events with energies above $10^{19}$\,eV.  We scan from 
$0^{\circ}$ to $5^{\circ}$ in steps of $0.1^{\circ}$ in angular
separation.  The energy threshold $E_{min}$ is lowered one event at 
a time, starting at the highest energy event and decrementing to
$10^{19}$\,eV.

For the HiRes stereo data set, the strongest potential signal
is found at $E_{min}=1.69\times 10^{19}$\,eV and 
$\theta_{min}=2.2^{\circ}$.  The chance probability for this 
potential signal is $P_{min}=1.9\,\%$, but the final chance 
probability after accounting for the scan is $P_{ch}=52\,\%$.  
The signal is therefore not significant, and we conclude
that the HiRes stereo data above $10^{19}$\,eV is consistent 
with the null hypothesis of isotropic arrival directions.  As a 
consequence of our approach, we find that this conclusion is 
indeed very general.  There is no evidence for significant clustering 
for any energy threshold above $10^{19}$\,eV on all angular scales 
of $5^{\circ}$ or less.  This indicates that clustering of arrival 
directions is weaker than previously suggested and might not be a 
general feature of ultrahigh energy cosmic rays after all.

\section{Comparison to the AGASA Data Set}

The HiRes stereo data set is still smaller than the published AGASA 
data set.  In the energy range of interest, above 
$4\times 10^{19}$\,eV, it has 
significantly less statistical power than the AGASA data set, but the 
difference in the number of events stems largely from a difference in 
observed flux rather than from a large difference in exposure.  The 
discrepancy in flux is larger than the statistical errors of the 
experiments, and the possibility of a systematic energy shift of order 
$30\,\%$ as studied in~\cite{demarco2003} implies 
that by using the same energy threshold for both experiments, we are 
not comparing equivalent data sets.  Our scan over energies is partly 
motivated by this energy discrepancy.  If the HiRes counterpart to the 
AGASA energy threshold is somewhat lower than $4\times 10^{19}$\,eV, then 
the scan at least ensures that we have tested for small-scale clustering 
in a HiRes data set equivalent in energy threshold to the published AGASA 
set.

In order to evaluate the sensitivity of the current HiRes stereo data 
set to ultrahigh energy cosmic ray point sources, we can set upper flux 
limits on sky locations which might harbor sources.  The flux limits 
will be a function of the equatorial coordinates of any potential source, 
since the detector exposure is not uniform.  As an example, we determine 
upper flux limits for proton sources of constant intensity at the locations 
of the AGASA multiplets.

To do this we track the ``source'' on its path over the (local) sky for 
every day of data taking between December 1999 and January 2004, and
simulate a constant rate of events with a differential energy spectrum 
$\propto E^{-2.71}$ from these positions using the correct zenith 
and azimuth angles.  We use an hourly database of running times, weather
conditions, and detector parameters to generate events.   After 
reconstruction, the quality cuts applied to the real data are also 
applied to the simulated events.  We determine the number of events 
reconstructed within $2.5^{\circ}$ of the ``source'' direction
and compare this number to the generated flux to calculate a $90\,\%$ 
upper flux limit using~\cite{helene1983}.
Table\,\ref{table} shows the result for the positions of the 5 AGASA
multiplets above $4\times 10^{19}$\,eV.

\begin{table}[htb]
\caption{\it HiRes stereo flux upper limits for the positions of the 
AGASA clusters.}
\label{table}
\renewcommand{\arraystretch}{1.2} 
\begin{tabular}{@{}crrr}
\hline
    &\multicolumn{2}{c}{equatorial coord.} & $F^{90\%}\,(> 40\,\mathrm{EeV}) $ \\
Cluster    & $\alpha~~[^{\circ}]$ & $\delta~~[^{\circ}]$ & 
          $[\mathrm{km}^{-2}\,\mathrm{yr}^{-1}]$ \\
\hline
1  &  18.6  &  20.6  & 0.023 \\
2  & 169.7  &  56.9  & 0.016 \\
3  & 283.0  &  48.1  & 0.026 \\
4  &  63.9  &  30.0  & 0.021 \\
6  & 213.1  &  37.4  & 0.021 \\
\hline
\end{tabular}\\[2pt]
AGASA cluster numbers according to ref. \cite{agasa1999}.
\end{table}

For AGASA, we calculate a flux from the locations of the
clusters by using the exposure published for the data set
until May 2000.  For each of the doublets
(clusters 1, 3, 4, and 6 in Table\,\ref{table}), the flux is 
about $0.010\,\mathrm{km}^{-2}\,\mathrm{yr}^{-1}$, while for the
triplet (cluster 2 in Table \,\ref{table}), the flux is
$0.013\,\mathrm{km}^{-2}\,\mathrm{yr}^{-1}$.  The HiRes stereo
flux upper limits can at this point not rule out the flux 
estimated from the AGASA clusters.  Nevertheless, for some 
regions of the sky, the HiRes stereo exposure comes to within 
about $25\,\%$ of the exposure of the published AGASA data set.

\section{The AGASA Clustering Signal}

To understand the extent of the inconsistency between the HiRes
result and AGASA's claim that ultrahigh energy cosmic ray arrival 
directions exhibit significant small-scale clustering, we must better 
understand the AGASA clustering signal itself.  An unbiased test
of the clustering hypothesis is only possible if the data set used
for formulating the hypothesis is not also used for evaluating 
the significance.  Crucial parameters defining this signal,
{\it i.e.} the energy threshold $4\times 10^{19}$\,eV and the 
maximum angular distance between cluster events of $2.5^{\circ}$,
were first described by the AGASA collaboration in~\cite{agasa1996}.  
Based on data recorded through October 1995, these cuts are justified 
{\it a posteriori} as cuts that lead to a significant clustering signal.  
Consequently, the strength of this signal should be evaluated using only 
data recorded since then.  This split in ``original'' and ``new'' data 
was introduced in~\cite{finley2004}, and the chance probability of the
clustering signal in the new data set based on the cuts defined
by the original set was found to be $27\,\%$.  If one allows 
for cross-correlations with the original set while replacing its
clusters by single events at their averaged position, the statistical
power of the new set can be increased, but the chance probability
for clustering in the new set is still $8\,\%$, implying that the
clustering signal in the ``new'' set is consistent with the null
hypothesis.  

One can take this independent test one step further.
With 27 events, the new data set is slightly smaller than the
original data set with 30 events.  If we extend the new data set
by including the data taken since May 2000, summarized
in a skymap on the AGASA web page~\cite{agasaweb}, the original
claim can now be tested with a larger data set comprising 42 events.
The chance probability for the clustering observed in the enlarged 
new set is $19\,\%$, so the larger data set fails to confirm the 
clustering hypothesis formulated using the original data set.

These tests show that the persistent claims that the clustering 
signal has chance probabilities as low as $10^{-6}$~\cite{agasa2001} 
are a direct consequence of the bias carried over from the original 
data set if the pre- and post-October 1995 data are not clearly 
separated.

With evidence of small-scale clustering in the AGASA data set being 
weak to insignificant, our result is therefore consistent with 
AGASA's.  It appears that at present, both data sets show no 
clustering signal.

HiRes is currently the only cosmic ray detector probing the 
northern sky. The statistical power of the HiRes data set is
increasing, and the important topic of small-scale anisotropies
in the arrival directions of cosmic rays at the highest energies
will be revisited once more data becomes available.

\vskip1cm

We thank the organizers of CRIS2004 for an exciting conference and for
their hospitality.
The HiRes project is supported by the National Science Foundation under
contract numbers NSF-PHY-9321949, NSF-PHY-9322298, NSF-PHY-9974537,
NSF-PHY-0098826, NSF-PHY-0245428, by the Department
of Energy Grant FG03-92ER40732, and by the Australian Research Council.
The cooperation of Colonels E. Fisher and G. Harter, the US Army and
Dugway Proving Ground staff is appreciated.  
We thank the authors of CORSIKA for providing us with
the simulation code.


\begin{thebibliography}{9}
\bibitem{sigl2003} G. Sigl, F. Miniati, and T. Ensslin, Phys. Rev. D 68 (2003) 044008.
\bibitem{dolag2003} K. Dolag, D. Grasso, V. Springel, and I.I. Tkachev,
   JETP Lett. 79 (2004) 583.
\bibitem{agasa1996} N. Hayashida et al., Phys. Rev. Lett.77 (1996) 1000.
\bibitem{agasa1999} M. Takeda et al., Astrophys. J. 522 (1999) 225.
\bibitem{agasa2001} M. Takeda et al., Proc. 27th ICRC, Hamburg, Germany (2001) 345.
\bibitem{tinyakov2001} P.G. Tinyakov and I.I. Tkachev, JETP Lett. 74 (2001) 1.
\bibitem{agasa2003} M. Teshima et al., Proc. 28th ICRC, Tsukuba, Japan (2003) 437.
\bibitem{watson2001} A.A. Watson, Proc. XIII Rencontres de Blois (2001), also
   arXiv: astro-ph/0112474.
\bibitem{evans2003} N.W. Evans, F. Ferrer, and S. Sarkar, 
   Phys. Rev. D 67 (2003) 103005.
\bibitem{finley2004} C.B. Finley and S. Westerhoff, 
   Astroparticle Phys. 21 (2004) 359.
\bibitem{thomson2004} G.B. Thomson et al., these proceedings.
\bibitem{corsika1998}
   D. Heck et al., CORSIKA: A Monte Carlo Code to
   Simulate Extensive Air Showers, Forschungszentrum
   Karlsruhe, Wissenschaftliche Berichte FZKA 6019 (1998).
\bibitem{apjl2004} R.U. Abbasi et al. (HiRes Collaboration), 
   Astrophys. J. 610 (2004) L73.
\bibitem{sadowski2002} P.A. Sadowski et al. (HiRes Collaboration), 
   Astroparticle Phys. 18 (2002) 237.
\bibitem{demarco2003} D. De Marco, P. Blasi, and A.V. Olinto,
   Astroparticle Phys. 20 (2003) 53.
\bibitem{helene1983} O. Helene, Nucl. Instr. Meth. 212 (1983) 319.
\bibitem{agasaweb} http://www-akeno.icrr.u-tokyo.ac.jp/AGASA/results.html.
\end{thebibliography}
\end{document}